# Scalable Compression of a Weighted Graph

Kifayat Ullah Khan, Waqas Nawaz∗, Young-Koo Lee

Data and Knowledge Engineering Lab, Department of Computer Science and Engineering, Kyung Hee University, 446-701, Republic of Korea

∗Institute of Information Systems, Innopolis University, Universitetskaya St. 1, Innopolis, Tatarstan Republic, Russia 420500.

{kualizai, yklee}@khu.ac.kr, ∗w.nawaz@innopolis.ru

## Abstract

Graph is a useful data structure to model various real life aspects like email communications, co-authorship among researchers, interactions among chemical compounds, and so on. Supporting such real life interactions produce a knowledge rich massive repository of data. However, efficiently understanding underlying trends and patterns is hard due to large size of the graph. Therefore, this paper presents a scalable compression solution to compute summary of a weighted graph. All the aforementioned interactions from various domains are represented as edge weights in a graph. Therefore, creating a summary graph while considering this vital aspect is necessary to learn insights of different communication patterns. By experimenting the proposed method on two real world and publically available datasets against a state of the art technique, we obtain order of magnitude performance gain and better summarization accuracy.

## 1. INTRODUCTION

Use of web and social media is common for daily life activities. People from all walks of life rigorously utilize these online services to fulfill their needs. Therefore, underlying graph is a rich and large repository due to real world interactions. However, efficiently finding useful knowledge from a graph is hard, due to its massive size. We find that clustering a graph helps to learn its overall charactertics [5]. However, it does not reduce size of a graph for efficient execution of various graph mining operations. On the other hand, summary graph is useful as it is smaller in size and maintains properties of its underlying graph.

We observe that many real world graphs are weighted. Edge weights are useful to show various aspects like number of co-authored papers, emails exchanged between persons, and so on. To create summary of a weighted graph, Toivonen et al. [3] propose a pioneering research. The authors present various algorithms to create an insightful summary graph. Core of their algorithms is to create summary graph by utilizing pair-wise nodes merging strategy like [2]. We find that this super node creation strategy is in-efficient to create summary of large graph. The reason is that every iteration performs similarity checkings to find best pair of nodes for compression, which are at-least equal to 2-hop neighborhood of each node. However, only a pair of nodes are compressed after performing such large number of computations. As a result of this, summary generation requires large amount of time.

To improve execution time of pair-wise summarization method, we present a Set-based Graph Summarization (SAGS) methodology in our previous work to create summary of a non-weighted graph [1]. This set-based strategy provides efficiency, since a set of nodes are aggregated in each iteration. Hence, a summary graph is computed in short time. To identify set of nodes for compression, we utilize a well-known similarity search technique, Locality Sensitive Hashing (LSH). LSH is an approximate technique for fast nearest neighbor search in high dimensional data. In this paper, we extend our set-based merging from [1] for compression of a weighted graph.

## 2. PROBLEM DEFINITION

Our objective in this paper is to efficiently compute a concise summary $S_G$ for a given large graph $G$. We require $S_G$ to be computed in-time and concise so that it can support memory-based execution of various graph mining algorithms.

To efficiently compute $S_G$, we aggregate sets of nodes in each iteration during summary graph creation by using SAGS.

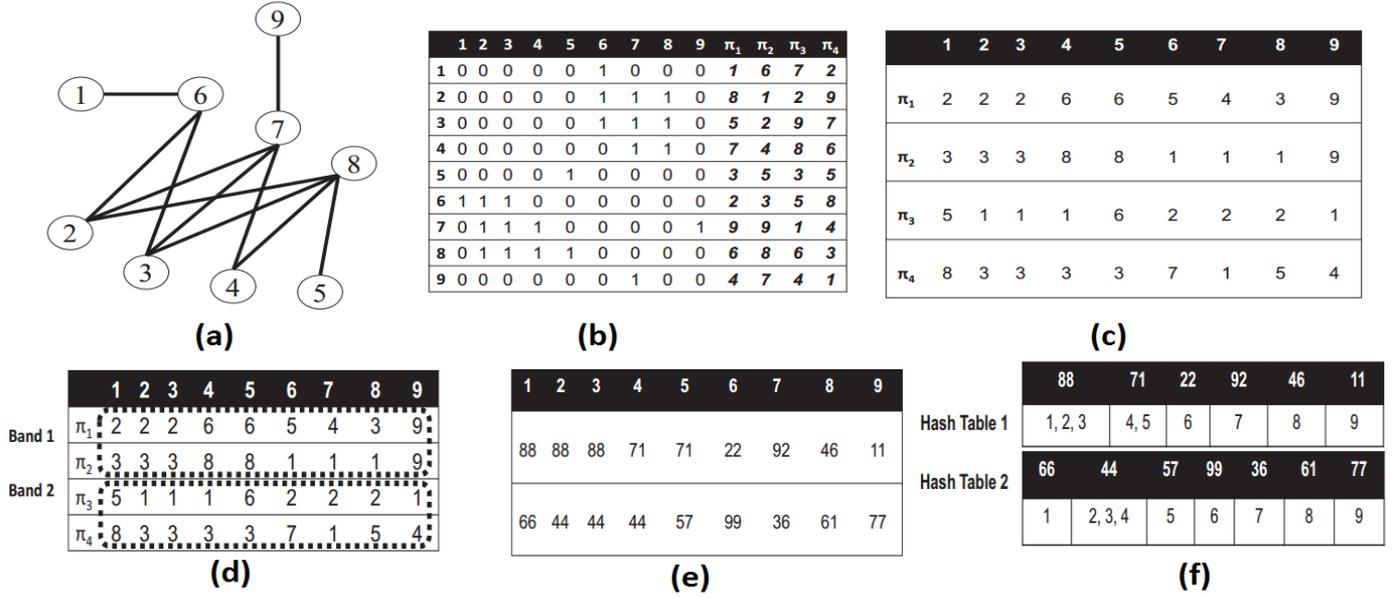

**Fig. 1**: Illustration of applying LSH on a simple graph [1]. (a) Sample Graph (b) Adjacency matrix representation of sample graph, along with four hash functions of random permutations based on node ids (c) Minhash matrix (d) Division of matrix into 2 bands (e) Combined hash codes for each portion of columns of every band (f) Hash tables containing buckets of candidate similar nodes.

Similarly, we control the conciseness of $S_G$ using compression ratio $cr$ in Equation (1) from [3]. In case of edge weights, we set weight of each super edge as the mean weight from its member edges to minimize the error.

$$\text{cr}(S_G) = |E_S|/|E| \qquad (1)$$

**Problem Statement:** Given an undirected and weighted graph $G(V, E, w)$ and compression ratio $cr$, $0 < cr < 1$. our objective is to compute its $S_G(V_S, E_S, w_S)$ in linear time, where $cr(S_G) \leq cr$, and $dist\big(G, dcomp(S_G)\big)$ is minimal.

## 3. METHODOLOGY

In this section, we present our methodology to create summary of a weighted graph. We first explain our set-based merging method and then detail how to generate candidate nodes using LSH.

**Set-based Merging**: Set-based super node creation strategy aggregates set of nodes in each iteration to create an $S_G$. This set-based approach is efficient since merging multiple nodes, reduce total number of graph traversal iterations. To identify a set of nodes for compression, we select a query node $q$, where $q_i = v_i \in V$ to reduce all pairwise similarity computations and to find mergable nodes for compression.

**Generating Candidate Similar Nodes:** The step-by-step processing for hashing candidate similar nodes from a simple (non-weighted) graph is illustrated in Fig. 1 [1]. We find that this hashing scheme does not consider weights attached to edges. Therefore, we first generate candidate nodes like those from a simple graph, and then directly filter out highly similar nodes by using post-pruning for edge weights. In this way, we prune every node that violates thresholds for compression using neighborhood similarity as well as edge weight difference.

**Filtering Compressible Nodes:** To filter out highly similar nodes for a given $q$, we first retrieve its candidate similar nodes, from all the buckets where it exists. After retrieval, we sort them based on their cost reduction [2] against $q$ and then iteratively prune every candidate node that violates thresholds of neighborhood similarity and edge weight difference. In this way, resultant set is is compressed into a super node. This process is repeated until entire graph is visited.

## 4. EXPERIMENTS

We perform experiments on two real world weighted graphs. Our summary evaluation criteria is execution time and Root Mean Square Error (RMSE) due to difference of edge weights between decompressed version of summary and original graph. We perform comparisons against state of the art compression solution (Randomized algorithm) for a weighted graph [3]. *bRandomized* is modified version of Randomized where number of super nodes in summary graph is same as that by SAGS. Similarly, *iSAGS* includes indexing time to generate candidate similar nodes. We find that execution of SAGS is order of magnitude faster than its competitor and RMSE is also better. Missing results for RMSE indicate that SAGS cannot generate summary graph at these compression ratios. The reason is that Randomized only considers similarity of edge weights for each merger. So it keeps on merging every pair of node until required compression ratio is met. On the other hand, SAGS considers both neighborhood as well as edge weight similarity, hence, number of compressible sets for compression are lesser in count.

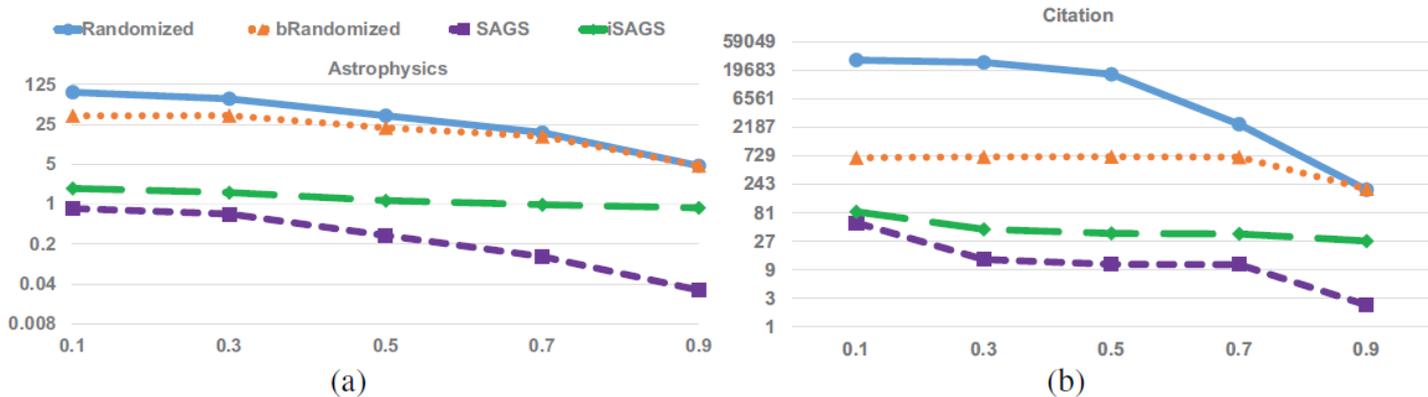

Fig. 2: Execution Time Comparison at different Compression Ratios. X and Y axis respectively show compression ratio and execution time in seconds.

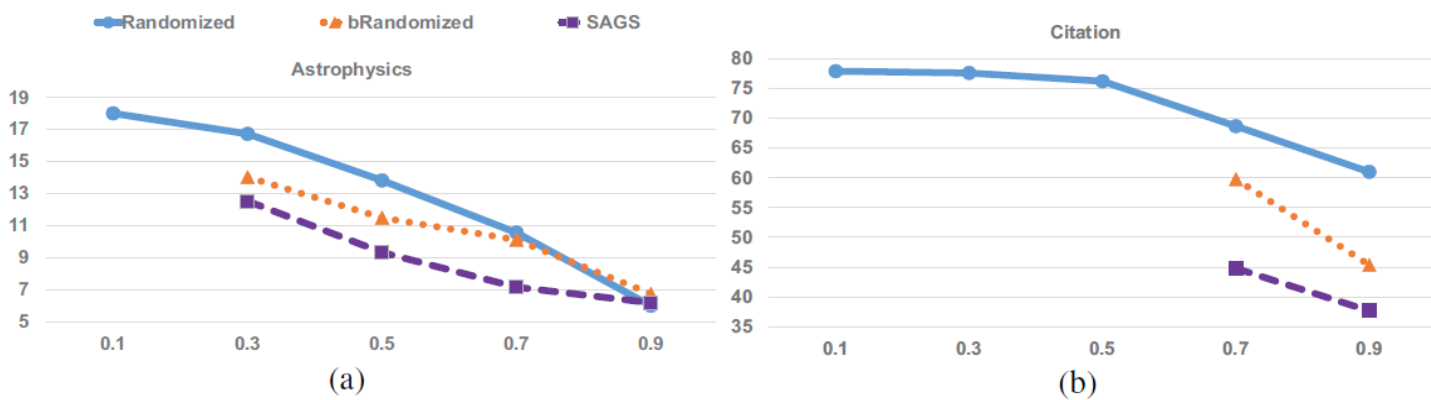

Fig. 3: RMSE at different Compression Ratios. X and Y axis show compression ratio and RMSE respectively.

## 5. CONCLUSION

We present fast summarization of a large weighted graph in this paper. Summary of a weighted graph is necessary to learn insights of relationships among its member nodes. In future, our objective is to extend our work for summary computation of a weighted and dynamic graph.

## ACKNOWLEDGEMENT

This work was supported by the National Research Foundation of Korea (NRF), funded by the Korean Government (MEST) (No.2015R1A2A2A01008209).